\documentclass[prl,showpacs,twocolumn,superscriptaddress,floatfix,amsmath]{revtex4}
\usepackage[english]{babel}
\usepackage{graphicx}
\usepackage{amsmath}
\usepackage{amssymb}

\begin{document}
\title{Electrostatic confinement of electrons in an integrable graphene quantum dot}

\author{J.\ H.\ Bardarson}
\affiliation{Laboratory of Atomic and Solid State Physics, Cornell
  University, Ithaca, NY 14853, USA}
\author{M.\ Titov}
\affiliation{School of Engineering \& Physical Sciences, Heriot-Watt
  University, Edinburgh EH14 4AS, UK}
\author{P.\ W.\ Brouwer}
\affiliation{Laboratory of Atomic and Solid State Physics, Cornell
  University, Ithaca, NY 14853, USA}
\date{\today}
\begin{abstract}
We compare the conductance of an undoped graphene sheet with a
small region subject to an electrostatic gate potential for the
cases that the dynamics in the gated region is regular 
(disc-shaped region) and classically chaotic (stadium). For the disc, we find 
sharp resonances that narrow upon reducing the area fraction of the 
gated region. We relate this observation to the existence of 
confined electronic states. For the stadium, the 
conductance looses its dependence on the gate voltage upon reducing 
the area fraction of the gated region, which signals the lack of
confinement of Dirac quasiparticles in a gated region with chaotic
classical electron dynamics.
\end{abstract}


\maketitle{}

\begin{figure}[tb!]
  \begin{center}
    \includegraphics[width=0.80\columnwidth]{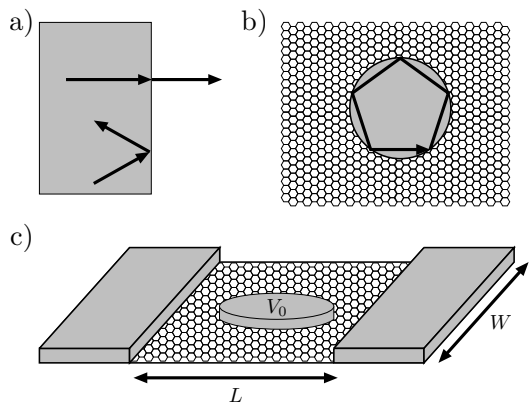}
  \end{center}
  \caption{Angle-dependent scattering of an electrostatic interface (a), a circular quantum dot 
  (defined by an electrostatic gate potential) in an infinite graphene sheet showing an example of
  a path that corresponds to a bound states (b), and a two terminal transport setup that can probe for the existence of bound
  states (c).}
  \label{fig:setup}
\end{figure}

A characteristic feature of the massless Dirac electrons in graphene,
a single layer of graphite, is the suppression of backscattering
\cite{And98} and the resulting strong angle dependence of scattering at
interfaces \cite{Che06} (related to ``Klein tunneling''
\cite{Kat06,Bee08}). For clean graphene, graphene without
atomic-scale defects or impurities that cause intervalley scattering, 
the transmission
probability through an electrostatic barrier is unity at perpendicular
incidence and drops sharply upon increasing the angle of incidence 
(cf.\ Fig.~\ref{fig:setup}a). The absence of backscattering at 
perpendicular incidence is a consequence of a combination of an 
effective time reversal symmetry and band
topology~\cite{And98,Ryu07}. The same combination is responsible for
the fact that, generically, Dirac electrons can not be confined 
in electrostatically defined structures or by a smooth random electrostatic
potential through Anderson localization \cite{Nom07,Bar07}. For 
this reason, most
experimental ~\cite{Pon08, Sta08} and theoretical \cite{Wur09, Lib09}
studies of confined electrons in graphene are performed on etched
structures, in which inevitably intervalley scattering is strong and 
the physics essentially reduces to that of normal semiconductor quantum
dots.

This general argument against the possibility to confine electrons
electrostatically  overlooks a special case: Dirac
electrons can be confined electrostatically if their classical
dynamics inside the gated area is
integrable and the corresponding Dirac equation is separable. 
The integrable classical dynamics allows for the existence
of electron paths for which the electron never approaches the
boundary of the confinement area 
at perpendicular incidence. For such paths, an electrostatic
barrier is fully reflecting, so that the electrons are effectively
confined. In quantum mechanics, it is the corresponding separability 
of the Dirac equation that invalidates the topological arguments 
against the electrostatic confinement of electrons \cite{Nom07,Fu06}.

The requirement of integrable dynamics is essential for the
electrostatic confinement of the electrons. Certainly, if the 
classical dynamics 
is chaotic, electrons can not be confined electrostatically. There is
a rich history of examples in which the question of integrability of
the classical dynamics has profound consequences for a quantum 
mechanical system. The best known example is that of the level 
spacing distribution of ballistic normal-metal quantum dots 
\cite{Bee97,Alh00}, which is described by Wigner-Dyson statistics for
quantum dots with chaotic classical dynamics, and Poisson statistics
for quantum dots with integrable classical dynamics. Another example, 
similar in spirit to the present study, is the lifetime of resonant
states in optical resonators or microlasers
\cite{Yam93,Joh93}. Regular lasers have optical ray trajectories that 
always exceed the critical angle and thus have an infinite lifetime, 
while chaotic trajectories are always eventually refracted out and
thus have a finite lifetime~\cite{Noc97}. 

Experimentally, the existence of bound states would be clearly
revealed in a two terminal conductance measurement setup schematically
shown in Fig.~\ref{fig:setup}c. A ``buffer'' of undoped graphene separates
a gated area from the leads and acts as an effective barrier. (The
barrier is only ``effective'' because the conductivity of 
undoped graphene is $4
e^2/\pi h$, not zero.) A metal gate defines a ``quantum dot'' in 
which the carrier density is nonzero.
If the dot is located approximately halfway
between the metal contacts, eventual bound states in the dot will
appear as resonances in the conductance upon varying the gate voltage
of the dot. Since only regular dots can support bound states, the
presence of sharp resonances in the two terminal
conductance gives a clear experimental signature of integrability in
the graphene quantum dot.

The remainder of this letter is reserved for material in support of the
above scenario. We begin by considering, as in Fig.~\ref{fig:setup}b, a circular quantum dot in an 
infinite graphene sheet. This problem can be solved exactly, and the
location and character of the bound states can be identified
explicitly. We then turn to the two-terminal transport geometry of
Fig.\ \ref{fig:setup}c and
compare numerical simulations of transport through a layer of undoped
graphene with circular and stadium-shaped quantum dots

Taking the gate potential to be smooth on the scale of the lattice 
spacing, the electron dynamics in graphene is accurately described by 
the single valley Dirac Hamiltonian  
\begin{equation}
  H_0 = v_{\rm F}\mathbf{p}\cdot\mathbf{\sigma} + V_{\rm gate}(x,y)
  \label{eq:H}
\end{equation}
where $\mathbf{\sigma} = (\sigma_x,\sigma_y)$ are Pauli matrices and $v_{\rm F}$ is the
Fermi velocity. For a circular dot we take the gate potential to be
\begin{equation}
  V_{\rm gate}(r) = \hbar v_{\rm F}V_0\vartheta(r-R),
  \label{eq:gate}
\end{equation}
with $\vartheta$ the Heaviside step function. (The choice of a step
function allows a closed-form solution; it is not essential for the
existence of bound states. The structure of quasi-bound states in the inverted setup [zero potential inside, nonzero outside] was considered in
Ref.~\onlinecite{Mat08}.) We look for eigenstates of $H_0$ at zero 
energy. Because of the rotation symmetry of $V_{\rm gate}$, the
eigenstates can be labeled according to their angular momentum $m$,
with $m$ half-integral~\cite{Rec07}. They are
\begin{equation}
  \psi_m 
    = e^{im\theta}\begin{pmatrix} e^{-i\theta/2}\varphi_{m,+}(r) \\ e^{i\theta/2}\varphi_{m,-}(r) \end{pmatrix},
  \label{eq:full}
\end{equation}
where the radial functions $\varphi_{m,\sigma}$ with $\sigma = \pm$ satisfy
  \begin{equation}
    -i\left[ \partial_r -\sigma (m-\frac{\sigma}{2})\frac{1}{r}\right]\varphi_{m,\sigma} = -V_0\vartheta(R-r)\varphi_{m,-\sigma}. 
  \label{eq:radial}
\end{equation}
For $r > R$ the two equations~\eqref{eq:radial} decouple and the solutions are 
\begin{equation}
  \varphi_{m,\sigma} = a_\sigma r^{\sigma m-1/2}. \\
  \label{eq:outside}
\end{equation}
Normalizability requires $|m| \geq 3/2$ and $a_+=0$ ($a_-=0$) for $m>0$ ($m<0$). 
For $r < R$ one finds 
\begin{subequations}
  \begin{align}
    \varphi_{m,+}(r) &= bJ_{|m-1/2|}(|V_0|r), \\
    \varphi_{m,-}(r) &= -ib\,{\rm sgn}(V_0m)J_{|m+1/2|}(|V_0|r).
  \end{align}%
\label{eq:inside}%
\end{subequations}%
Continuity of the wave function at $r=R$ then gives the
implicit equation  
\begin{equation}
  J_{|m|-1/2}(|V_0|R) = 0,
  \label{eq:zeros}
\end{equation}
from which one obtains the values of the gate voltage $V_0$ for which bound states exist at zero energy. These states correspond to states with nonzero
orbital angular momentum that circulate inside the disc, thereby avoiding Klein tunneling. The states are twofold degenerate,
corresponding to clockwise and anti-clockwise circulation. This
twofold degeneracy is not specific to the circular shape of the
quantum dot, but
is rather an example of the Kramers degeneracy due to the effective time reversal, $\mathcal{T} = i\sigma_y\mathcal{C}$ with
$\mathcal{C}$ complex conjugation, of the single valley Dirac
Hamiltonian~\eqref{eq:H}. 

The bound states exist for $|m| \ge 3/2$ only. At angular momentum
$|m| = 1/2$ no normalizable bound state exists. Instead, the ``zero
mode'' at $|m|=1/2$ is an extended state. It is the existence of
this extended zero mode that is responsible for the fact that 
electrons can not be confined electrostatically in a chaotic dot: For
a chaotic dot, all wavefunctions will in general have a finite 
contribution from the zero mode, which will determine the finite 
lifetime of the state. This is to be contrasted with the case of the
circle studied above, which admits solutions to the Dirac equation in 
which the zero mode has vanishing weight. 

\begin{figure}[tb!]
  \begin{center}
    \includegraphics[width=0.80\columnwidth]{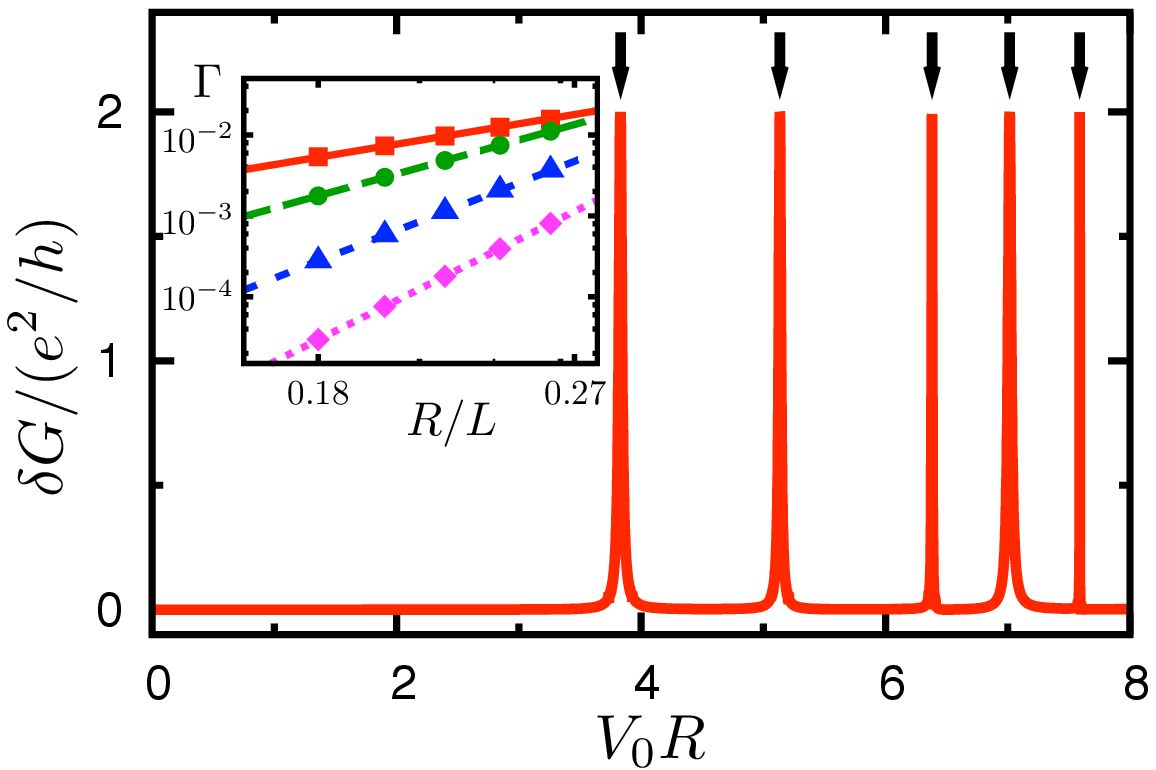}\\
    \vspace{0.2cm}
     \includegraphics[width=0.80\columnwidth]{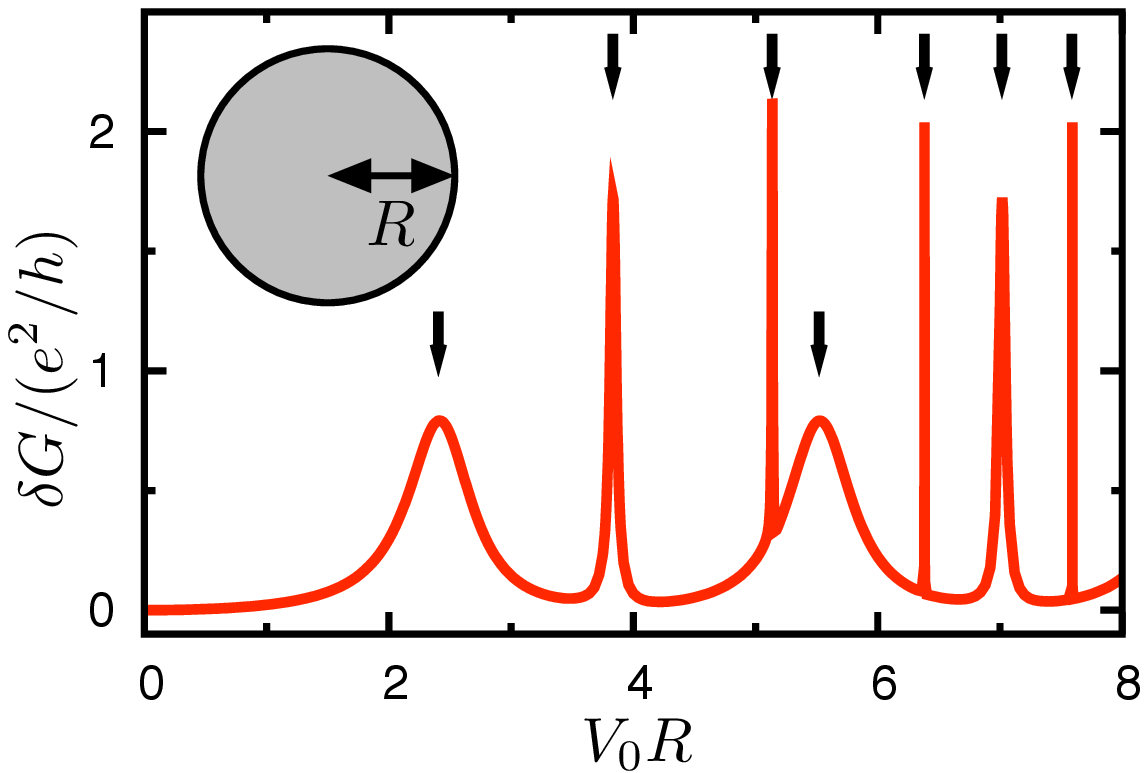}

  \end{center}
\caption{Conductance of an undoped graphene sheet with
an electrostatically gated circular region of radius $R$ (lower inset)
placed in 
the center of the sheet, as a function of the gate voltage $V_{0}$ 
for $R/L = 0.26$, $W/L = 1$ (upper panel) and $R/L = 0.2$, $W/L = 6$. The arrows show the positions of the solution to 
Eq.~\eqref{eq:zeros}. Upper inset the dependence of the width of the first few resonances on dot size $R/L$.}
  \label{fig:full}
\end{figure}

Following Ref.~\onlinecite{Two06}, transport properties in the two-terminal 
geometry of Fig.\ \ref{fig:setup}c are described by the Hamiltonian
\begin{equation}
  H = H_0 + U_{\rm lead},
  \label{eq:Htrans}
\end{equation}
where $U_{\rm lead}(x) = 0$ for $0 < x < L$ and $U_{\rm lead}(x)
\rightarrow \infty$ otherwise. The graphene sheet has width $W$.
The conductance is
obtained numerically using the transfer matrix method of 
Refs.~\onlinecite{Tit07} and~\onlinecite{Bar07}. We place the center
of the quantum dot at $x = L/2$. 
In Fig.~\ref{fig:full} we show the results of such a calculation for
the conductance for an undoped graphene sheet with a circular quantum dot,
as a function of the gate voltage $V_0$ for aspect ratios $W/L=1$ and
$W/L=6$ and periodic boundary conditions in the transverse ($y$)
direction. In both cases, we observe sharp resonances on top of a background 
conductance from the finite conductivity of the undoped graphene buffer.
The positions of the sharp resonances are in good agreement with the gate
voltages~\eqref{eq:zeros} of the bound states.
We have also performed the same calculation with antiperiodic boundary 
conditions (results not shown) and obtained results that are
qualitatively similar to those for periodic boundary conditions if
$W/L=1$ and indistinguishable from these if $W/L=6$ (as was to be
expected \cite{Two06}).

For aspect ratio $W/L=1$, the background conductance is essentially
independent of the gate voltage $V_0$. For large aspect ratios,
however, the background conductance has a relatively slow gate-voltage
dependence from a contribution from the zero mode. This is a finite 
size effect. However, the density of states associated with the zero mode 
goes to zero only logarithmically in the limit $L/R \gg 1$, which
is why this feature persists for the entire range of system sizes
accessible numerically. Still, the sharp resonances from the true 
bound states can be unambiguously identified, and their positions 
are independent of aspect ratio and boundary conditions.

The width of the resonances in Fig.~\ref{fig:full} depends on the
angular momentum quantum number $m$ of the resonances. The dependence
on $m$ and on the size $L$ of the undoped graphene buffer separating
the dot from the electrodes can be understood in a simpler model that
retains the circular symmetry. To that end we replace the leads by a
circular electrode at a distance $L$ from the center of the quantum
dot. Since the integrated probability of the bound states at distance
$L$ from the origin scales as $(R/L)^{2 m}$, we conclude that the
resonance width $\Gamma$ should have the same dependence on $L$ and
$m$,
\begin{equation}
\Gamma \sim (R/L)^{2|m|}. 
  \label{eq:Gamma}
\end{equation} 
The larger the angular momentum quantum number $m$ the narrower the resonances become.

\begin{figure}[tb!]
  \begin{center}
    \includegraphics[width=0.80\columnwidth]{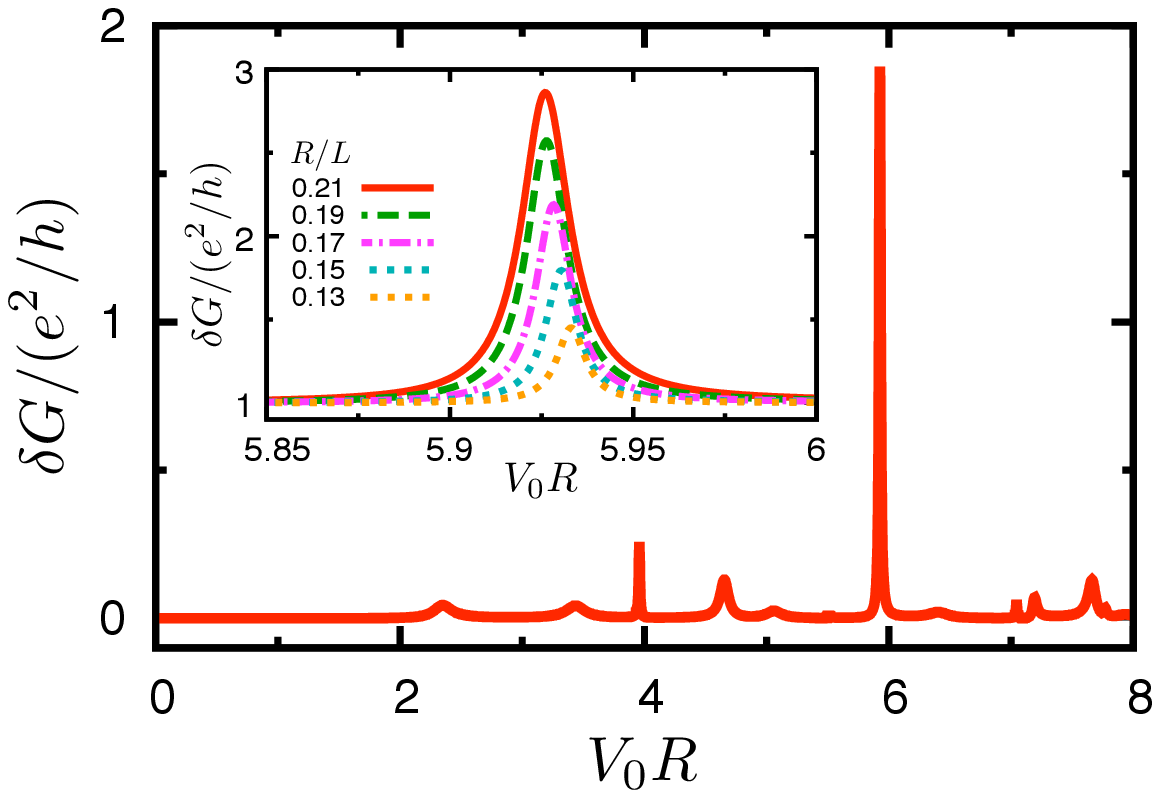}\\
    \includegraphics[width=0.80\columnwidth]{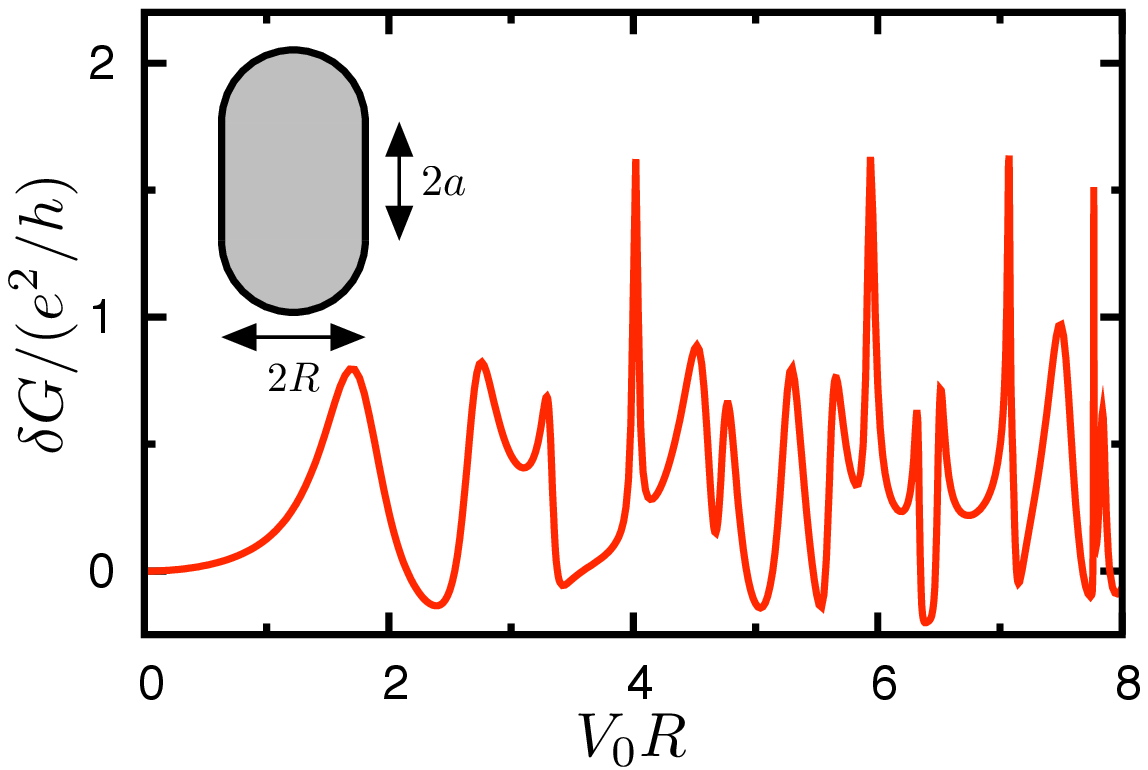}
  \end{center}
\caption{Same as Fig.\ \ref{fig:full}, but with a
stadium-shaped gated region (lower inset) with $2 a/R =
\sqrt{3}$ and $R/L = 0.21$, $W/L = 1$ (upper panel) and $R/L = 0.2$, $W/L = 6$ (lower panel). The dependence of the sharpest
resonance in the upper panel on the ratio $R/L$ is shown in the upper inset.}
  \label{fig:StadFull}
\end{figure}

We have verified the prediction~\eqref{eq:Gamma} for the resonance
width by repeating the numerical calculation for different values of
$R/L$. 
The resonances have a Lorentzian shape, and we extract their width by 
fitting to the formula
\begin{equation}
  \delta G = \frac{2\Gamma^2}{(V_0R - x_0)^2 + \Gamma^2}
  \frac{e^2}{h},
  \label{eq:Lorentzian}
\end{equation}
where $x_0$ is approximately given by the zero of the Bessel 
function~\eqref{eq:zeros}. (There is a small shift of the resonance
position from the coupling to the leads.) The dependence of the
resonance width on $R/L$ shown in the upper inset to Fig.~\ref{fig:StadFull} is in excellent agreement with the 
prediction (\ref{eq:Gamma}) independent of boundary conditions and aspect ratio.

Other regular shapes that give rise to integrable classical dynamics 
and a separable Dirac equation behave in the same way as the circle. 
One analytically solvable example is given by a strip of width  
$R$ extending across the entire width of the sample. For $W \simeq L$ the conductance has separated Lorentzian resonances as a
function of gate voltage, comparable to the circular gate voltage of Fig~\ref{fig:full}, reflecting discrete bound states. In the
limit $ R \ll L \ll W$ these bounds states become effectively
degenerate and their resonances start to overlap. The resonances are
at position $V_0R=\pi(n+1/2)$ with integer $n$ and their height diverges (in the limit $U_{\rm lead}\to \infty$),
\begin{equation}
G= \frac{e^2}{h}\frac{W}{\pi L} 
\frac{1}{\sin V_0R}\ln\left| \frac{1+\sin V_0R}{\cos V_0R}\right|.
\end{equation}

As a prototypical example of a chaotic quantum dot we consider the
conductance through a stadium (see lower inset in
Fig.~\ref{fig:StadFull}). We have given arguments for why the stadium
can not support bound states and we thus do not expect to see any
sharp resonances in the conductance. This is confirmed by the
calculated conductance trace in Fig.~\ref{fig:StadFull}, which shows
wide features which have a height that decreases and a width that
remains constant upon increasing $L/R$ (provided $L/R$ is large
enough). This is shown explicitly in
the inset to the upper panel of Fig.~\ref{fig:StadFull} for the sharpest
conductance peak near $V_0 R = 6$. Just as in the circular case contributions from quasibound states appear for larger aspect
ratios. The behavior of these features is very similar to the zero mode contributions in the circle.

\begin{figure}[tb!]
  \begin{center}
    \includegraphics[width=0.80\columnwidth]{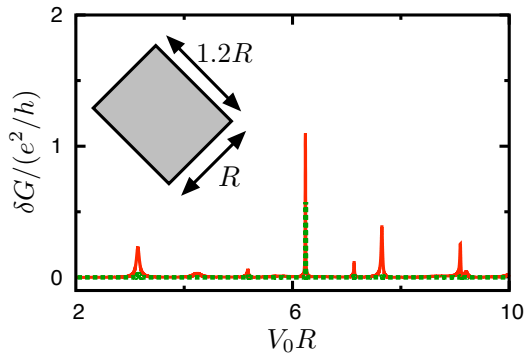}
  \end{center}
  \caption{Same as Fig.~\ref{fig:full}, but with an
  an electrostatically gated rectangular region of size $R \times 1.2
  R$ and rotated by 45 degrees relative to the leads placed in the
  center of the sheet. The solid and dotted curves correspond 
to $R/L = 0.2$ and
  $0.1$, respectively. \label{fig:5}}
\end{figure}

For the Dirac equation, integrability of the classical equation of motion does not 
always imply integrability of the quantum mechanical equation of 
motion. An example is a uniformly 
gated region with a rectangular shape, 
for which the classical dynamics is integrable, but the Dirac equation
is not~\cite{Ber87}. For the gate voltage range we considered in our
numerical calculations (up to $\sim 50$ levels in a circle of
comparable size), we found no conductance resonances that have the
same behavior upon increasing the ratio $L/R$ as resonances in a 
circular quantum dot (becoming narrower at constant height); 
in contrast all
resonances in the range of gate potentials we considered become 
lower upon increasing $L/R$. (Typical data are
shown in Fig.~\ref{fig:5}.) 
This implies that, for this gate
voltage range, a necessary condition for confinement is
separability of the quantum mechanical equation of motion, not
merely integrability of the classical dynamics. 
We can not numerically determine whether sharp resonances and, 
hence, confined electronic states, emerge at higher gate voltages, 
where the quantum-classical correspondence plays a more significant
role.

In conclusion, we have shown that gate potentials in which the
(quantum) equation of motion is integrable, surrounded by undoped
graphene, allow electrons to be confined. Gated regions in which the
classical dynamics is chaotic, on the other hand, do not support  
bound states. Bound states manifest
themselves in a two terminal conductance measurement through the
presence of sharp resonances. Compared to the complexity of
experiments on etched quantum dots in graphene, the setup suggested 
here is a promising candidate for the observation of
signatures of integrable dynamics in graphene.

We are grateful to C.\ Kane for a stimulating discussion.
This work was supported by the NSF under Grant No.\ DMR 0705476.

\end{document}